\documentclass[aps,prd,nofootinbib,twocolumn,showpacs]{revtex4}
\usepackage{dcolumn}
\usepackage{lipsum}
\usepackage{graphicx,epsfig}
\usepackage{psfrag}
\usepackage{bm}
\usepackage{epstopdf}
\usepackage{latexsym}
\usepackage{multirow}
\usepackage{amsmath,amssymb}
\usepackage{enumerate}
\usepackage{hyperref}
\usepackage[FIGTOPCAP]{subfigure}

\def\be {\begin{equation}}
\def\ee {\end{equation}}
\def\ba {\begin{eqnarray}}
\def\ea {\end{eqnarray}}
\def\nn {\nonumber}
\def\bc {\begin{center}}
\def\ec {\end{center}}
\newcommand{\bdm}{\begin{displaymath}}
\newcommand{\edm}{\end{displaymath}}

\def\a  {\alpha}
\def\b  {\beta}
\def\g  {\gamma}
\def\G  {\Gamma}
\def\d  {\delta}

\def\l  {\lambda}
\def\m  {\mu}
\def\n  {\nu}
\def\o  {\omega}
\def\O  {\Omega}

\def\r  {\rho}

\def\s {\sigma}

\def\t  {\tau}

\def\nn {\nonumber}
\def\ra {\rightarrow}
\def\na {\nabla}
\def\da {\dagger}
\def\la {\label}
\def\le {\left}
\def\ri {\right}
\def\pa {\partial}
\def\f {\frac}
\def\sq {\sqrt}

\def\bi {\begin{itemize}}
\def\ei {\end{itemize}}

\def\> {\rangle}
\def\< {\langle}

\def\bc {\begin{center}}
\def\ec {\end{center}}

\begin{document}

\title{Creation of spin $1/2$ particles and renormalization in FLRW spacetime} 

\author{Suman Ghosh} \email[email: ]{smnphy@gmail.com}
\affiliation{Department of Theoretical Sciences, S N Bose National Centre for Basic Sciences, Kolkata - 700098, India}

%

\begin{abstract}
Within the framework of adiabatic regularization
, we present a simple formalism to calculate number density and renormalized energy-momentum density of spin 1/2 particles in spatially flat FLRW spacetimes using an appropriate WKB ansatz for the adiabatic expansion for the field modes. 
The conformal and axial anomalies thus found are in exact agreement with those obtained from other renormalization methods. This formalism can be considered as an appropriate extension of the techniques originally introduced for scalar fields, applicable to fermions in curved space.
\end{abstract}

\maketitle


\section{Introduction}

Quantum field theory in curved space-time \cite{PT,MW,Wald-QFTCS,Fulling,BD} has been developed as an approximate quantum theory of gravity in order to study particle creation by evolving universe \cite{Parker-thesis, Parker:1968mv, Parker:1969au, Parker:1971pt} and  black holes \cite{Hawking:1974sw}, as well as inhomogeneities in the cosmic microwave background radiation and the large-scale structure of the Universe \cite{Liddle:2000cg}. Parker \cite{Parker-thesis} conceptualised the so-called adiabatic vacuum in order to have a notion of particles in curved space that comes closest to the usual one in flat space. 
For scalar fields in spatially flat Friedmann-Lemaitre-Robertson-Walker (FLRW) spacetimes -- (i) in a co-moving volume, the particle number density ($|\b_k|^2$) is an adiabatic invariant, 
(ii) particles of conformally invariant field with zero mass will not be created, (iii) the total number density of created particles of specific mass, summed over all modes is ultraviolet (UV) divergent and (iv) the stress tensor ($T_{\m\n}$) of created particles has quadratic and logarithmic UV  divergences in addition to the expected quartic divergence. 
%
%

Various renormalization methods were developed to tame these infinities. 
The concept of {\em adiabatic regularization} was introduced by Parker \cite{parker:2012} to make the total particle number density for scalar particles finite and later extended to tame the UV divergences in $T_{\m\n}$  by Parker and Fulling \cite{Parker:1974qw}.
In adiabatic regularization, the physically relevant finite expression is obtained from the formal one containing UV divergences by subtracting mode by mode (under the integral sign) each term in the adiabatic expansion of the integrand that contains at least one UV divergent part for arbitrary values of the parameters of the theory. The number of time derivatives of the cosmological scale factor $a(t)$ that appear in a term of the expansion is called the {\em adiabatic order} of the term. The adiabatic regularization scheme is particularly useful for numerical calculations.
In \cite{Parker:1974qw}, the authors have also shown that the adiabatic regularization is equivalent to the n-wave regularization (which is essentially a variant of Pauli-Villars regularization \cite{BD}) used by Zeldovich and Starobinsky \cite{Zeldovich:1971mw} to renormalize the divergent $T_{\m\n}$ for scalar fields in an anisotropic universe.
Like adiabatic regularization scheme, this method is also particularly suited for numerical computations \cite{Garriga:1989jx, Ghosh:2008zs}. 

Among other standard techniques, proper-time regularization, point-splitting regularization (particularly by Hadamard method), zeta-function regularization and dimensional regularization have been applied to curved space \cite{PT,MW,Wald-QFTCS,Fulling,BD}. 
The DeWitt-Schwinger point-splitting regularization \cite{DeWitt:1975ys, Christensen:1976vb, Birrell-1978, Anderson:1987yt, Christensen:1978yd}, is recently used in \cite{Matyjasek:2013vwa, Matyjasek:2014lja} to construct an approximate $T_{\m\n}$ of the quantized massive scalar, spinor and vector fields in the spatially flat FLRW universe using asymptotic expansion of the Green function constructed within the framework of the n-wave regularization \cite{Zeldovich:1971mw, Beilin:1980es} and reproduced the leading-order contribution to the stress tensor derived in \cite{Kaya:2011yu}. All these methods are equivalent and leads to the same output \cite{BD}. 
Production of spin $1/2$ particles in various cosmological scenarios have been studied by many \cite{Unruh:1974bw, Campos:1991ff, Gibbons:1993hg, Villalba:1995za, Brustein:2000hi, Anischenko:2009va}.
Recently fundamental issues like problem of defining a preferred vacuum state at a given time is addressed in \cite{Agullo:2014ica, Anderson:2005hi}.
Analysis of the approximate definition of the particle number via an adiabatic WKB ansatz can be found in \cite{Winitzki:2005rw}.
A systematic adiabatic expansion for spin $1/2$ modes has been recently constructed in \cite{Landete:2013axa, Landete:2013lpa, delRio:2014cha} to analyze $|\b_k|^2$ for fermions and the corresponding renormalization of $T_{\m\n}$ in FLRW universe and is used to prove the equivalence between  adiabatic regularization and point-splitting DeWitt-Schwinger renormalization \cite{delRio:2014bpa}. It was argued in \cite{Landete:2013axa} that WKB ansatz is specifically designed to preserve the Klein-Gordon product and the associated Wronskian condition, but not to preserve the Dirac product and the normalization condition. 


In the following, we present a simple formalism to determine $|\b_k|^2$ and also to regularize the resulting $T_{\m\n}$ for spin $1/2$ particles in spatially flat FLRW universe.
The essential difference between our formalism and that introduced in \cite{Landete:2013axa} is explicit in the corresponding expressions of socalled `out' states. In our case, the entire non-adiabaticity is incorporated in the Bogolubov coefficients whereas in \cite{Landete:2013axa}, the Bogolubov coefficients are defined to be of particular adiabatic order. 
Further, we have expressed quantum fields and other quantities as functions conformal time which is useful particularly in conformally flat spacetimes and lead us to simple structure of field equations. 
We define our `in' state as the adiabatic vacuum given by the WKB solution to the field equations and the `out' state as mixture of positive and negative frequency `in' states via {\em time-dependent} Bogolubov coefficients $\a_k(t)$ and $\b_k(t)$. Next we use the field equations to derive the governing equations for $|\b_k|^2$, in terms of a set of three {\em real} and independent variables $s_k, u_k$ and $\t_k$\footnote{Similarly, in \cite{Landete:2013axa}, one has to solve for three complex quantities, namely, $\o$, $F$ and $G$.} 
(to be defined later), that were introduced in \cite{Zeldovich:1971mw} in the context of scalar particle creation during anisotropic collapse. 
It is then straightforward to regularize $T_{\m\n}$ by subtracting leading order terms from the adiabatic  mode expansions  of these variables. The renormalized quantities thus derived match exactly with the known results found by other methods \cite{BD, PT, Landete:2013axa}. 
%

\section{Dirac field in FLRW spacetime}  

The homogeneous and isotropic FLRW spacetime geometry is given by
\be
ds^2 = a^2(t)  (- dt^2 + d\vec{x}^2), \la{eq:metric}
\ee
where $t$ is the conformal time and $a(t)$ is the conformal scale factor. The Dirac equation in generic curved spacetime for field $\psi(\vec{x},t)$ with mass $m$ is given by \cite{PT,MW,Wald-QFTCS,Fulling,BD},
\be
(e^\m_a\g^a \nabla_\m - m)\psi = 0. \la{eq:Dirac1}
\ee
where $e^a_\m$ are the vierbeins, $\g^a$'s are standard Dirac matrices (defined in terms of usual Pauli matrices $\s^i$)  in Minkowski space, satisfying $\{\g^a,\g^b\} = \eta^{ab}$ and $\nabla_\m = \pa_\m - \G_\m$ is the covariant derivative with $\G_\m$ being the spin connection \cite{PT,BD}. 
The Dirac matrices (compatible with signature $-,+,+,+$) in the Dirac-Pauli representation is given by
\ba
\g^0 = i \le( \begin{array}{cc} 
I & 0 \\
0 & -I \\ 
\end{array} \ri), ~~~~~ \g^i = i\le( \begin{array}{cc} 
0 & \s^i   \\
-\s^i & 0 \\ 
\end{array} \ri)
\ea
where $\s^i$ are the usual Pauli matrices given by
\ba
\s^1 = \le( \begin{array}{cc} 
0 & 1   \\
1 & 0 \\ 
\end{array} \ri),~~\s^2 = \le( \begin{array}{cc} 
0 & -i   \\
i & 0 \\ 
\end{array} \ri),~~\s^3 = \le( \begin{array}{cc} 
1 & 0   \\
0 & -1 \\ 
\end{array} \ri).
\ea
For metric (\ref{eq:metric}), Eq. (\ref{eq:Dirac1}) leads to
\be
\le[\g^0 \le(\pa_0 + \f{3\dot{a}}{2a}\ri) + \g^i \pa_i + ma\ri] \psi = 0. \la{eq:Dirac2}
\ee	
Dirac field $\psi$ can be written in terms of a {\em time-dependent} annihilation operator for particles ($B_{\vec{k}\l}(t)$) and creation operator for antiparticles ($D^{\da}_{\vec{k}\l}(t)$) as
\be
\psi = \sum_{\l} \int d^3k \big(B_{\vec{k}\l}u_{\vec{k}\l} + D^{\da}_{\vec{k}\l}v_{\vec{k}\l}\big) \la{eq:psi-gen}
\ee
where momentum expansion of the eigenfunctions $u_{\vec{k}\l}(\vec{x},t)$ and $v_{\vec{k}\l}(\vec{x},t)$, which is obtained by charge conjugation ($v = \g^2u^*$) operation on $u_{\vec{k}\l}(\vec{x},t)$, in terms of two component spinors are given by \cite{Landete:2013axa},
\ba
u_{\vec{k}\l}(\vec{x},t) &=& \f{e^{{i \vec{k}.\vec{x}}}}{(2\pi a)^{3/2}} \le( \begin{array}{l} 
h^{I}_{k}(t) \xi_\l(k)  \\
h^{II}_{k}(t) \f{\vec{\s}.\vec{k}}{k} \xi_\l(k) \\ 
\end{array} \ri)  \la{eq:+psi_ansatz} \\
v_{\vec{k}\l}(\vec{x},t) &=& \f{e^{{-i \vec{k}.\vec{x}}}}{(2\pi a)^{3/2}} \le( \begin{array}{l} 
-h^{II*}_{k}(t) \f{\vec{\s}.\vec{k}}{k} \xi_{-\l}(k)  \\
-h^{I*}_{k}(t)  \xi_{-\l}(k) \\ 
\end{array} \ri)  \la{eq:-psi_ansatz}
\ea
where $\xi_\l(k)$ is the normalised two-component spinor satisfying $\xi_\l^\da \xi_\l = 1$ and $\f{\vec{\s}.\vec{k}}{2k}\xi_\l = \l\xi_\l$ where $\l = \pm 1/2$ represents the helicity\footnote{Note that using either value of helicity or either of $u_{\vec{k}\l}$ and $v_{\vec{k}\l}$, leads to exactly same end results in the following calculations.}. Normalisation condition in terms of Dirac product for $\psi$, $(u_{\vec{k}\l}, u_{\vec{k'}\l'}) = (v_{\vec{k}\l}, v_{\vec{k'}\l'}) = \d_{\l\l'}\d(\vec{k}-\vec{k'})$, implies
\be
|h^{I}_{k}(t)|^2 + |h^{II}_{k}(t)|^2 = 1. \la{eq:norm_h}
\ee
This condition guaranties the standard anti-commutation relations for creation and annihilation operators.
Putting Eq. (\ref{eq:+psi_ansatz}) in Eq. (\ref{eq:Dirac2}) we get the following first order coupled differential equations
\ba 
{\dot h}^{I}_k + i m a\, h^{I}_{k} + i k\, h^{II}_{k} &=& 0,  \la{eq:h1h2}\\
{\dot h}^{II}_k - i m a\, h^{II}_{k} + i k\, h^{I}_{k} &=& 0. \la{eq:h2h1}
\ea
and the Wronskian is given by
\be
{\dot h}^{I}_k\, h^{II*}_k - h^{I}_k\,{\dot h}^{II*}_k = -i k. \la{eq:wronskian1} 
\ee
Here ($\dot{}$) means derivative with respect to the conformal time `$t$'. Eq. (\ref{eq:h1h2}) and (\ref{eq:h2h1}) leads to the following decoupled second order equations
\ba
{\ddot h}^I_k + \big[\O_k^2(t) + i Q(t)\big] h^{I}_{k} &=& 0, \la{eq:h1} \\
{\ddot h}^{II}_k + \big[\O_k^2(t) - i Q(t)\big] h^{II}_{k} &=& 0, \la{eq:h2}
\ea
where $\O_k(t) = \sq{m^2a^2 + k^2}$ and $Q(t) = m \dot{a}$. Note that, the following methodology is applicable to generic backgrounds where equations of similar structure as Eqs (\ref{eq:h1}) and (\ref{eq:h2}) appear.
The `in' adiabatic vacuum (i.e. state of adiabatic order zero) is given by the WKB solution (that naturally generalises the standard Minkowski space solution) of the field equations (see Appendix \ref{app-1}),
\ba
h^{I(0)}_k(t)  = f_1\, e_-,~~~~ h^{II(0)}_k(t)  = f_2\, e_- \la{eq:g1g2}
\ea
with $f_1 = \sq{\f{\O_k + ma}{2\O_k}},~~ f_2 = \sq{\f{\O_k - ma}{2\O_k}}$ and $e_{\pm} = \exp(\pm i\int \O_k dt).$
This implies, we shall seek the general solution corresponding to the `out' state in the form 
\ba
h^{I}_k(t) &=& \a_k(t) h^{I(0)} - \b_k(t) h^{II(0)*}, \la{eq:h1ansatz}\\
h^{II}_k(t) &=& \a_k(t) h^{II(0)} + \b_k(t) h^{I(0)*} \la{eq:h2ansatz}
\ea
where $\a_k(t)$ and $\b_k(t)$ are the Bogolubov coefficients. Eqs. (\ref{eq:h1ansatz}) and (\ref{eq:h2ansatz}) further implies 
\ba
\a_k(t) &=& \Big(f_1\, h^{I}_k + f_2\, h^{II}_k\Big)e_+,  \la{eq:alpha1} \\
\b_k(t) &=& \Big(f_1\, h^{II}_k - f_2\, h^{I}_k\Big)e_-. \la{eq:beta1}
\ea 
The normalisation condition (\ref{eq:norm_h}) leads to
\be
|\a_k(t)|^2 + |\b_k(t)|^2 = 1. \la{eq:norm_ab}
\ee
Then the average number of spin $1/2$ particles of specific helicity and charge with momentum $\vec{k}$ created per unit volume is given by \cite{PT},
\be
\langle N_{\vec{k}}\rangle = \langle B^{\da}_{\vec{k}\l} B_{\vec{k}\l}\rangle = \langle D^{\da}_{\vec{k}\l} D_{\vec{k}\l}\rangle = |\b_k(t)|^2. \la{eq:no.den.}
\ee
Note that, the WKB solutions (\ref{eq:g1g2}) obey the following Wronskian condition
\be
{\dot h}^{I(0)}_k\, h^{II(0)*}_k - h^{I(0)}_k\,{\dot h}^{II(0)*}_k = F - i k, ~~~~F = \f{k\,Q}{2\O_k^2}. \la{eq:wronskian2} 
\ee
The function $F(t)$ is of adiabatic order one and contains the factor $Q(t)$ that breaks the conformal invariance in the field equations. So the Wronskian is satisfied in the {\em adiabatic limit} and $F(t)$ is a measure of {\em non-adiabaticity} of the cosmological evolution.
Eq. (\ref{eq:wronskian2}) also implies that the WKB ansatz is not an exact solution of the field equation during the {\em non-adiabatic} expansion which is the desired condition to have any particle creation \cite{parker:2012}.
In Eq. (\ref{eq:psi-gen}), the creation and annihilation operators carry this non-adiabaticity and so do the Bogolubov coefficients in Eqs (\ref{eq:h1ansatz}-\ref{eq:h2ansatz}). Thus $|\b_k|^2$ is expected to depend on $F(t)$ as particle creation can be considered as a result of this non-adiabaticity.
Putting Eqs (\ref{eq:h1ansatz}) and (\ref{eq:h2ansatz}) in Eqs (\ref{eq:h1h2}) and (\ref{eq:h2h1}) and simplifying, a system of two linear first order differential equations is obtained for $\a_k(t)$ and $\b_k(t)$:
\be
\dot{\a}_k = -F\, \b_k\, e_+^2, ~~~~~~\dot{\b}_k = F\, \a_k\, e_-^2, \la{eq:alphabeta}
\ee
which were first derived by Parker \cite{Parker:1971pt} exactly in this particular form. It is obvious from Eq. (\ref{eq:alphabeta}) that creation of massless particles in conformally flat spacetimes is prohibited.
It also says that fermions at rest shall not be created. Similar results were found in \cite{Khanal:2013cjp}, where the authors have used Newman-Penrose formalism.

To determine $|\b_k|^2$ and the resulting $T_{\m\n}$, let us define the following three {\em real} and independent variables \cite{Zeldovich:1971mw} (for later convenience), in terms of the two complex variables $\a_k$ and $\b_k$ which are related by condition (\ref{eq:norm_ab}),
\ba
s_k = |\b_k |^2,~~ u_k = \a_k\,\b_k^*\,e_-^2 + \a_k^*\,\b_k\,e_+^2, \nn \\
\t_k = i(\a_k\,\b_k^*\,e_-^2 - \a_k^*\,\b_k\,e_+^2). \la{eq:def_sut}  
\ea
For these variables one gets a system of three linear first order differential equations:
\ba
\dot{s}_k &=& F\ u_k, \la{eq:s}\\
\dot{u}_k &=& 2F(1-2s_k) - 2 \O_k \t_k, \la{eq:u}\\
\dot{\t}_k &=& 2\O_k u_k. \la{eq:t}
\ea
with initial conditions $s_k = u_k = \t_k = 0$ at some suitably chosen $t= t_0$. 

From equations (\ref{eq:s}-\ref{eq:t}) one can further decouple the equation for $|\b_k|^2$ or $s_k$ given as
\be
\dddot{s}_k + F_1 \ddot{s}_k + F_2 \dot{s}_k + F_3 (1-2s_k) = 0, \la{eq:s_3} 
\ee
where 
\ba
F_1 &=& - \le(\f{2\dot{F}}{F} + \f{\dot{\O}_k}{\O_k}\ri), \la{eq:F1}\\
F_2 &=& \le(4F^2 + 4\O_k^2 + \f{2\dot{F}^2}{F} + \f{\dot{F}\dot{\O}_k}{F\O_k} - \f{\ddot{F}}{F}\ri), \la{eq:F2}\\
F_3 &=& - 2F^2 \le(\f{\dot{F}}{F} - \f{\dot{\O}_k}{\O_k}\ri). \la{eq:F3}
\ea
Note that once $h^{I}_k$ and $h^{II}_k$ are derived (analytically or numerically) from the field equations, one can find $|\b_k|^2$ directly from Eq. (\ref{eq:beta1}). Alternatively, when such closed form solution to the field equations are not available, one can solve the set of equations (\ref{eq:s}-\ref{eq:t}) numerically (or analytically whenever possible). 
Below we discuss how one can find the renormalized stress tensor by solving Eqs (\ref{eq:s}-\ref{eq:t}).

\subsection{Energy-momentum tensor}

The energy-momentum tensor for Dirac field in curved spacetime is given by
\be
T_{\m\n} = \f{i}{2} \Big[\bar{\psi}\g_{(\m}\na_{\n)}\psi - (\na_{(\m}\bar{\psi})\g_{\n)}\psi\Big] \la{eq:stress}
\ee
The independent components of $T_{\m\n}$ are given by
\ba
T^0_0 &=& -\f{i}{2a} \Big(\bar{\psi}\g^0\dot{\psi} - \dot{\bar{\psi}}\g^0\psi\Big), \la{eq:T00}\\
T^i_i &=& \f{i}{2a} \Big(\bar{\psi}\g^i\psi' - \bar{\psi}'\g^i\psi\Big), \la{eq:Tii}
\ea
where ($'$) denotes derivative with respect to $x^i$. Vacuum expectation value of the above quantities leads to
\ba
\< T^0_0 \>  &=& \f{1}{(2\pi a)^3} \int d^3k \, \r_k, \la{eq:<T00>}\\
\< T^i_i \> &=& \f{1}{(2\pi a)^3} \int d^3k \, p_k, \la{eq:<Tii>}
\ea
with energy density $\rho_k$ and pressure density $p_k$ are given, respectively as,
\ba
\r_k &=& \f{i}{a} \Big(h^{I}_k \dot{h}^{I*}_k + h^{II}_k \dot{h}^{II*}_k - h^{I*}_k \dot{h}^{I}_k - h^{II*}_k \dot{h}^{II}_k\Big), \la{eq:r_k} \\
p_k &=&  \f{2k}{3a} \Big(h^{I}_k h^{II*}_k + h^{I*}_k h^{II}_k\Big). \la{eq:p_k}
\ea
Using Eqs (\ref{eq:h1ansatz}), (\ref{eq:h2ansatz}), (\ref{eq:norm_ab}) and (\ref{eq:def_sut}), we get from Eqs (\ref{eq:r_k}) and (\ref{eq:p_k})
\ba
\r_k &=& -\f{2\O_k}{a}\big(1-2s_k\big), \la{eq:rho_k_sut} \\
p_k &=& \f{2k}{3a} \le[\f{k}{\O_k}(1- 2s_k) + \f{ma}{\O_k} u_k\ri]. \la{eq:p_k_sut}
\ea
The vacuum energy (when $s_k = u_k = \t_k = 0$) matches with the standard result. 
We discuss below how to remove the divergences in $T_{\m\n}$ by subtracting the leading order terms from the adiabatic expansion of $s_k$, $u_k$ and $\t_k$.


\subsection{Renormalization}

Let us consider the case of large momenta ($\O_k \ra \infty$) and expand the solutions of the system of Eqs (\ref{eq:s}-\ref{eq:t}) in an asymptotic series in powers of $\O_k^{-1}$. This is essentially same as the adiabatic expansion that is valid in the quasi-classical region where $|\dot{\O}_k|<< \O_k^2$. It is straightforward to see that $\t_k = \t_k^{(1)} + \t_k^{(3)} + ...$, $u_k = u_k^{(2)} + u_k^{(4)} + ...$ and $s_k = s_k^{(2)} + s_k^{(4)} + ...$, where the superscripts inside the brackets indicates the adiabatic order (Appendix \ref{app-2}). 
Eqs (\ref{eq:s}-\ref{eq:t}) leads to the following recursion relations 
\ba
u_k^{(r)} &=& \f{\dot{\t}_k^{(r-1)}}{2\O_k}, \la{eq:u^r}\\
s_k^{(r)} &=& \int F u_k^{(r)} dt, \la{eq:s^r}\\
\t_k^{(r+1)} &=& -\f{4F s_k^{(r)} + \dot{u}_k^{(r)}}{2\O_k}, \la{eq:t^r}
\ea
with $r = 2,4,...$ and $\t_k^{(1)} = \f{F}{\O_k}$. It is straightforward to solve these equations analytically to arbitrary order. Further, as $k \ra \infty$, we have
\be
s_k^{(r)} \sim k^{-(r+2)},~~ u_k^{(r)} \sim k^{-(r+1)}, ~~ \t_k^{(r)} \sim k^{-(r+1)}. \la{eq:lim_sut}
\ee
This implies the well-known logarithmic UV divergences of the total energy and pressure density. Note that, no quadratic divergence appears for fermions as it does for scalar fields \cite{PT}. To remove these infinities, we need to subtract leading terms upto second order from the expansion of $s_k$ and $u_k$. This prescription is thus equivalent to adiabatic regularization. 
The total particle number density of a specific mass with summed over momenta is simply given as
\be 
N_m = \f{1}{(2\pi a)^3} \int d^3k \,s_k. 
\la{eq:totalPND}
\ee
The renormalized total energy and momentum density (after subtracting the vacuum contribution i.e. the quartic divergence) are given by
\ba
{\< T^0_0 \> }_{ren} &=& \f{2}{\pi^2 a^4} \int dk\, k^2\, \O_k \Big(s_k - s_k^{(2)}\Big), \la{eq:r_k_ren} \\
{\< T^i_i \> }_{ren} &=&  \f{1}{3\pi^2 a^4} \int dk \f{k^3}{\O_k} \Big[-2k \big(s_k  - s_k^{(2)}\big) \nn \\
&& \hspace{1.5cm} + ma \big(u_k  - u_k^{(2)}\big)\Big]. \la{eq:p_k_ren}
\ea
Note that in more generic spacetimes the fourth-order adiabatic terms may give rise to proper UV divergences \cite{Christensen:1978yd} and the renormalized quantities become
\ba
{\< T^0_0 \> }_{ren} &=& \f{2}{\pi^2 a^4} \int dk\, k^2\, \O_k \Big(s_k - s_k^{(2)} - s_k^{(4)}\Big), \la{eq:r_k_ren_gen} \\
{\< T^i_i \> }_{ren} &=&  \f{1}{3\pi^2 a^4} \int dk \f{k^3}{\O_k} \Big[-2k \big(s_k  - s_k^{(2)} - s_k^{(4)}\big) \nn \\
&& \hspace{1.5cm} + ma \big(u_k  - u_k^{(2)} - u_k^{(4)}\big)\Big]. \la{eq:p_k_ren_gen}
\ea
According to standard approach of regularization one considers the fourth order adiabatic terms as {\em potentially} divergent \cite{PT,BD} and to compute the trace anomaly, Eqs (\ref{eq:r_k_ren_gen}) and (\ref{eq:p_k_ren_gen}) are used instead of Eqs (\ref{eq:r_k_ren}) and (\ref{eq:p_k_ren}).

\subsection{Conformal and axial anomalies}
The trace of the energy momentum tensor (\ref{eq:stress}) is $T^\m_\m = m \bar{\psi} \psi.$
Thus the trace vanishes for massless fields. However the renormalization procedure renders the quantum counterpart of $T^\m_\m$ finite. This phenomenon is known as conformal anomaly. 
The vacuum expectation value of the trace of stress tensor is given by
\be
\< T^\m_\m \> = \f{1}{(2\pi a)^3} \int d^3k \, \< T^\m_\m \> _k \la{eq:<Tmm>}
\ee
with
\ba
\< T^\m_\m \> _k &=&  -2m\Big(|h^{I}_k|^2 - |h^{II}_k|^2\Big) \la{eq:<Tmm>k} \\
&=& - 2m \le[\f{ma}{\O_k}(1-2s_k) - \f{k}{\O_k}u_k\ri] \la{eq:<Tmm>k_sut}
\ea
where we have again used  Eqs (\ref{eq:h1ansatz}), (\ref{eq:h2ansatz}), (\ref{eq:norm_ab}) and (\ref{eq:def_sut}). One can also derive Eq. (\ref{eq:<Tmm>k_sut}) using the identity $\< T^\m_\m \> _k = \r_k + 3p_k$.
In the limit $m\ra 0$, it is enough to subtract terms upto the second order (i.e. $s_k^{(2)}$ and $u_k^{(2)}$) to remove the UV divergence from $\< T^\m_\m \> $.
After subtracting the vacuum contribution, the resulting renormalized trace anomaly is given by,
\be
\< T^\m_\m \> _{ren} = \lim_{m\ra 0} \f{2m}{(2\pi a)^3} \int d^3k \le[\f{2ma}{\O_k} s_k^{(4)} + \f{k}{\O_k} u_k^{(4)}\ri]  \la{eq:<Tmm>ren1}
\ee
as only the fourth order term in the expansions of $s_k(t)$ and $u_k(t)$ survives in the $m\ra 0$ limit and in fact is independent of $m$.	
	Using explicit expressions of $s_k^{(4)}$ and $u_k^{(4)}$ in Eq. (\ref{eq:<Tmm>ren1}), we get
\be
\< T^\m_\m \> _{ren} = \f{11\dot{a}^4 - 29 a \dot{a}^2\ddot{a} + 12  a^2\dot{a}\dddot{a} + 9 a^2 \ddot{a}^2 - 3 a^3\ddddot{a}}{240\pi^2 a^8}. \la{eq:<Tmm>ren2}
\ee
To cross-check the above expression, note that the conformal anomaly can be expressed in terms of the curvature invariants by the following generic expression \cite{PT,BD}
\be
\< T^\m_\m \> _{ren} = \f{1}{(4\pi)^2} \le(A\, C_{\a\b\g\d}C^{\a\b\g\d} + B\, G + C\, \square R \ri), \la{eq:Tmm_gen}
\ee
where $C_{\a\b\g\d}$ is the Weyl tensor, $R$ is the Ricci scalar and $G$ is the Gauss-Bonnet invariant, given by $G = -2(R_{\a\b}R^{\a\b} - R^2/3)$ with $R_{\a\b}$ being the Ricci tensor. For conformally flat spacetimes (\ref{app-3}) Weyl tensor vanishes identically. Equating Eq. (\ref{eq:Tmm_gen}) with Eq. (\ref{eq:<Tmm>ren2}), we get $B = -11/360$ and $C=1/30$, which agrees with the known results \cite{PT,BD}. This proves the viability of the methodology presented here.

The classical axial current ($J^\m = \bar{\psi}\g^5\g^\m\psi$ where $\g^5 = i \g^1\g^2\g^3\g^4$) is conserved for a massless Dirac field. 
The quantum counterpart of the divergence of the axial current is given by
\ba
\< \na_\m j^\m \> &=& 2i m \< \bar{\psi}\g^5\psi \> \la{eq:axcur} \\
&=& -\f{4i m}{(2\pi a)^3} \int d^3k \Big(h^{I*}_kh^{II}_k - h^{I}_kh^{II*}_k\Big) \la{eq:axcur_h} \\
&=& \f{4 m}{(2\pi a)^3} \int d^3k \, \t_k. \la{eq:axcur_t}
\ea
This implies that the renormalized axial anomaly is given by,
\be
\< \na_\m j^\m \> _{ren} = \lim_{m\ra 0} \f{4 m}{(2\pi a)^3} \int d^3k \, \big(\t_k - \t_k^{(1)}\big). \la{eq:axan}
\ee
For $m\ra 0$, none of the terms in the right hand side of Eq. (\ref{eq:axan}) survives and the resulting axial anomaly vanishes as expected \cite{PT}.


\section{Summary}

We have constructed a simple formalism, within the framework developed in \cite{Parker-thesis, Zeldovich:1971mw, Parker:1974qw}, to compute number density and renormalized energy-momentum density of spin $1/2$ particles created during the evolution of spatially flat FLRW universes. We introduced appropriate WKB ansatz that satisfies the normalisation condition and the Wronskian condition upto the desired adiabatic order. The role of {\em non-adiabaticity} is crucial in defining the `out' vacuum. Here the Bogolubov coefficients carry all the adiabatic orders (so to speak), unlike \cite{Landete:2013axa}, where the Bogolubov coefficients are defined to be of some particular adiabatic order. 
We have expressed the physical quantities as simple linear combinations of three real and independent variables $s_k, u_k, \t_k$ which are defined in terms of the usual Bogolubov coefficients. Role of these variables is a distinguishing feature of the algorithm presented here and makes the process of renormalization simple. 
The evolution of these three variables are governed by three linear first order coupled differential equations. It is easy to solve these equations with appropriate boundary conditions. Further, using adiabatic approximation, one can find the adiabatic expansion of these variables in powers of momenta. Subtracting upto necessary leading-order terms from the expansion of $\< T_{\m\n} \> $, renormalization is achieved in the usual manner. The conformal and axial anomalies thus found are in exact agreement with those obtained from other renormalization methods that involve tedious calculations. 
To carry out all the steps, one need not solve the field equations analytically for the out vacuum and the whole process is suitable for numerical calculations too.
This work gives us a simple alternative to \cite{Landete:2013axa} as well as an appropriate extension and unification of standard techniques (within the framework of adiabatic regularization), originally introduced for scalar fields, applicable to fermions in curved space. Application of this formalism to interesting cosmological scenarios and the corresponding results will be reported elsewhere.

\section*{Acknowledgements}
The author thanks S. Kar, A. Lahiri and J. Navarro-salas for insightful discussions.

\appendix


\subsection{WKB solution}\la{app-1}
To find the WKB solution to 
\ba
{\ddot h}^I_k + \big[\O_k^2(t) + i Q(t)\big] h^{I}_{k} &=& 0, \la{eq:h1app} \\
{\ddot h}^{II}_k + \big[\O_k^2(t) - i Q(t)\big] h^{II}_{k} &=& 0, \la{eq:h2app}
\ea
let us assume
\be
h^{I}_k(t) \sim \exp\le[\int\Big(X(t) + i Y(t)\Big)dt\ri] \la{eq:wkbh1}
\ee
\be
\mbox{where}~~ X(t) = \f{1}{\hbar} \sum_{n=0}^{\infty}\hbar^n X_n(t), ~~Y(t) = \f{1}{\hbar} \sum_{n=0}^{\infty}\hbar^n Y_n(t). \la{eq:XY}
\ee
Putting Eq. (\ref{eq:wkbh1}) in Eq. (\ref{eq:h1app}) and equating the terms of zeroth order in $n$ we get
\ba
X_0^2 - Y_0^2 + \O_k^2 &=& 0, \la{eq:wkb11}\\ 
2X_0Y_0 + Q &=& 0. \la{eq:wkb12}
\ea
Similarly, solving for the first order in $n$, leads to
\ba 
\dot{X}_0 + 2X_0X_1 - 2Y_0Y_1 &=& 0, \la{eq:wkb21}\\ 
\dot{Y}_0 + 2X_0Y_1 - 2Y_0X_1 &=& 0. \la{eq:wkb22}
\ea
Higher order terms can be neglected in the adiabatic approximation. Solving Eqs (\ref{eq:wkb11}) and (\ref{eq:wkb12}) we get
\be
X_0 \approx \f{Q}{2\O_k}, ~~ Y_0 \approx \O_k.
\ee
Similarly, Eqs (\ref{eq:wkb21}) and (\ref{eq:wkb22}) gives
\be
X_1 \approx -\f{\dot{\O}_k}{2\O_k}, ~~ Y_1 \approx 0.
\ee
This leads to
\be
h^{I}_k(t) \sim \sq{\f{\O_k + ma}{2\O_k}} \exp\le[ i\int\O_k dt\ri]
\ee
Similarly one can solve Eq. (\ref{eq:h2app}). Note that the approximations made above are valid in the adiabatic limit. Therefore Eq. (\ref{eq:g1g2}) represents the adiabatic vacuum.


\subsection{$s_k$, $u_k$ and $\t_k$ of different adiabatic orders} \la{app-2}

Terms in adiabatic expansion of $s_k$, $u_k$ and $\t_k$ can be derived solving Eqs (\ref{eq:s}-\ref{eq:t}) in the following way. For $k\ra \infty$, $s_k$, $u_k$, $\t_k$ and there temporal variations must tend to zero. Therefore, Eq. (\ref{eq:u}) for large $k$ implies,
\be
0 \sim 2F - 2\O_k \t_k, 
\ee
which further implies that the leading term in the adiabatic expansion of $\t_k$ is of adiabatic order one, i.e.
\be
\t_k^{(1)} \sim \f{F}{\O_k} = \f{mk\dot{a}}{2\O_k^3}. \la{eq:t1} 
\ee
Putting Eq. (\ref{eq:t1}) in Eq. (\ref{eq:t}) we get the leading term in the adiabatic expansion of $u_k$ which is of order two, 
\be
u_k^{(2)} \sim \f{\dot{\t}_k^{(1)}}{2\O_k} = - \f{3m^3ka\dot{a}^2}{4\O_k^6} + \f{m k \ddot{a}}{4\O_k^4}.	 \la{eq:u2} 
\ee
Similarly, by putting Eq. (\ref{eq:u2}) in Eq. (\ref{eq:s}) we get the leading term in the adiabatic expansion of $s_k$ which is again of order two, 
\be
s_k^{(2)} \sim \int F u_k^{(2)} dt = \f{m^2k^2\dot{a}^2}{16\O_k^6}. \la{eq:s2} 
\ee
Now putting Eq. (\ref{eq:s2}) again back in Eq. (\ref{eq:u}) we get the next to leading term in the adiabatic expansion of $\t_k$ which is of adiabatic order three. This iteration leads to the equations (\ref{eq:u^r}-\ref{eq:t^r}) that give e.g.
\be
\t_k^{(3)} = \f{5m^3k^3\dot{a}^3}{16\O_k^9} - \f{15m^5ka^2\dot{a}^3}{8\O_k^9} + \f{5m^3k\dot{a}\ddot{a}}{4\O_k^7} - \f{mk\dddot{a}}{8\O_k^5}, \la{eq:t3} 
\ee

\ba
u_k^{(4)} &=& \f{35m^3k^5\dot{a}^2\ddot{a}}{32\O_k^{12}} + \f{5m^7ka^5\ddot{a}^2}{8\O_k^{12}} - \f{15m^7ka^5\dot{a}\dddot{a}}{16\O_k^{12}} + \f{5m^3k^5a\ddot{a}^2}{8\O_k^{12}}  \nn \\ 
&& - \f{105m^5k^3a\dot{a}^4}{32\O_k^{12}}  + \f{15m^3k^5a\dot{a}\dddot{a}}{16\O_k^{12}} + \f{105m^7ka^3\dot{a}^4}{16\O_k^{12}} \nn \\
&&  + \f{5m^5k^3a^3\ddot{a}^2}{4\O_k^{12}} + \f{15m^5k^3a^3\dot{a}\dddot{a}}{8\O_k^{12}} - \f{mk^7\ddddot{a}}{16\O_k^{12}} \nn \\ 
&& - \f{175m^5k^3a^2\dot{a}^2\ddot{a}}{32\O_k^{12}} - \f{3m^3k^3a^2\ddddot{a}}{16\O_k^{10}} - \f{105m^7k\dot{a}^2\ddot{a}}{16\O_k^{12}}, \la{eq:u4} 
\ea
\ba
s_k^{(4)} &=& \f{m^4k^4\dot{a}^4}{16\O_k^{12}} - \f{m^6k^2a^2\dot{a}^4}{4\O_k^{12}} + \f{7m^4k^2a\dot{a}^2\ddot{a}}{32\O_k^{10}} \nn \\
&& + \f{m^2k^2\ddot{a}^2}{64\O_k^8} - \f{m^2k^2\dot{a}\dddot{a}}{32\O_k^8}. \la{eq:s4}
\ea

Higher order terms can be derived in similar way. The Mathematica file containing these results are available on correspondence. This particular methodology introduced in \cite{Zeldovich:1971mw} for scalars has not been extended to deal with fermions as such. 




\subsection{Useful curvature quantities}\la{app-3}

Following are few useful formulas for FLRW geometry:
\be
R = 6\f{\ddot{a}}{a}, \la{eq:R}
\ee
\be
\square R = 6 \le(\f{3\ddot{a}^2}{a^6} - \f{6\dot{a}^2\ddot{a}}{a^7} + \f{4\dot{a}\dddot{a}}{a^6} - \f{\ddddot{a}}{a^5}\ri), \la{eq:boxR}
\ee
\be
R_{\m\n}R^{\m\n} = 12\le(\f{\dot{a}^4}{a^8} - \f{\dot{a}^2\ddot{a}}{a^7}	+ \f{\ddot{a}^2}{a^6}\ri). \la{eq:Rmnsq	}
\ee


\end{document}